\RequirePackage[l2tabu, orthodox]{nag}

\documentclass{sigchi}

\usepackage[T1]{fontenc}
\usepackage[utf8]{inputenc}
\usepackage[final,activate={true,nocompatibility},kerning=true,spacing=true]{microtype}
\usepackage[pdftex,hidelinks,unicode=true,pdfpagelabels=false]{hyperref}
\usepackage{balance}
\usepackage{listings}
\usepackage{siunitx}
\usepackage{graphicx}
\usepackage[table]{xcolor}
\usepackage{booktabs}
\usepackage{mathptmx}
\usepackage{xspace}
\usepackage{fixltx2e}
\usepackage{algorithm}
\usepackage[noend]{algpseudocode}
\usepackage{ntheorem}
\usepackage{amsmath}
\usepackage{amssymb}
\usepackage{paralist}
\usepackage{cleveref}

\crefrangelabelformat{line}{#3#1#4--#5#2#6}
\definecolor{kw}{rgb}{0.5,0.,0.33}
\definecolor{com}{rgb}{0.247,0.5,0.372}
\definecolor{str}{rgb}{0.165,0.0,1.0}
\definecolor{ano}{rgb}{0.4,0.4,0.4}
\definecolor{bg}{rgb}{1,1,1}

\lstdefinelanguage{MyJava}{
  stringstyle=\color{str}, 
  keywordstyle={\bfseries\color{kw}\ttfamily}, 
  commentstyle=\color{com}, 
  basicstyle={\scriptsize\ttfamily}, 
  captionpos=b,  
  frame=single,
  escapechar=|, 
  numbersep=5pt, 
  numbers=left, 
  language=java, 
  tabsize=2,
  backgroundcolor=\color{bg}, 
  moredelim=[il][\textcolor{ano}]{$$},
  moredelim=[is][\textcolor{ano}]{\%\%}{\%\%}
}

\DeclareGraphicsExtensions{.pdf, .png, .jpg}
\graphicspath{{res/}}

\makeatletter
\newcommand\footnoteref[1]{\protected@xdef\@thefnmark{\ref{#1}}\@footnotemark}
\makeatother

\newcommand{\tool}{\texttt{InspectorGuidget}\xspace}
\newcommand{\ie}{\emph{i.e.,}\xspace}
\newcommand{\eg}{\emph{e.g.,}\xspace}

\newcommand{\etal}{\emph{et al.}\xspace}
\newcommand{\etc}{\emph{etc.}\xspace}
\newcommand{\adhoc}{\emph{ad hoc}\xspace}

\newcommand{\change}[1]{#1}
\newcommand{\bl}{\emph{Blob listener}\xspace}
\newcommand{\bls}{\emph{Blob listeners}\xspace}

\newtheorem{definition}{Definition}

\def\pprw{8.5in}
\def\pprh{11in}

\lstnewenvironment{code}[1][]%
  {\noindent\minipage{\linewidth}\medskip 
   \lstset{#1}}
  {\endminipage}
  
\definecolor{linkColor}{RGB}{6,125,233}
\hypersetup{%
  pdftitle={Automatic Detection of GUI Design Smells: The Case of Blob Listener},
  pdfauthor={Valéria Lelli, Arnaud Blouin, Benoit Baudry, Fabien Coulon, Olivier Beaudoux},
  pdfkeywords={user interface, design smell, software validation, code quality},
  pdfdisplaydoctitle=true, 
  bookmarksnumbered,
  pdfstartview={FitH},
  colorlinks,
  citecolor=black,
  filecolor=black,
  linkcolor=black,
  urlcolor=linkColor,
  breaklinks=true,
  hypertexnames=false
}

\begin{document}

\toappear{Author's version.}

\title{Automatic Detection of GUI Design Smells:\\The Case of Blob Listener}


\numberofauthors{5}
\author{
\alignauthor
Val\'eria Lelli\\
\affaddr{University of Ceará, Brazil}\\
       valerialelli@great.ufc.br
\alignauthor
Arnaud Blouin\\
       \affaddr{INSA Rennes, France}\\
       arnaud.blouin@irisa.fr
\alignauthor
 Benoit Baudry\\
       \affaddr{Inria, France}\\
       benoit.baudry@inria.fr
\alignauthor
 Fabien Coulon\\
       \affaddr{Inria, France}\\
       fabien.coulon@inria.fr
\alignauthor Olivier Beaudoux\\
       \affaddr{ESEO, France}\\
       olivier.beaudoux@eseo.fr
}

 
\maketitle


\begin{abstract}
Graphical User Interfaces (GUIs) intensively rely on event-driven programming: 
widgets send GUI events, which capture users' interactions, to dedicated objects called \emph{controllers}.
Controllers implement several \emph{GUI listeners} that handle these events to produce GUI commands.
In this work, we conducted an empirical study on 13 large Java Swing open-source software systems.
We study to what extent the number of GUI commands that a GUI listener can produce has an impact on the change- and fault-proneness of the GUI listener code.
We identify a new type of design smell, called \bl that characterizes GUI listeners that can produce more than two GUI commands.
We show that \SI{21}{\percent} of the analyzed GUI controllers are \bls.
We propose a systematic static code analysis procedure that searches for \bl that we implement in \tool. 
We conducted experiments on six software systems for which we manually identified 37 instances of \bl.
\tool successfully detected 36 \bls out of 37.
The results exhibit a precision of \SI{97.37}{\percent} and a recall of \SI{97.59}{\percent}.
Finally, we propose coding practices to avoid the use of \bls.
\end{abstract}




\section{Introduction}\label{sec.intro}

Graphical User Interfaces (GUI) are the visible and tangible vector that enable users to interact with software systems.
While GUI design and qualitative assessment  is handled by GUI designers, integrating GUIs into software systems remains a software engineering task.
Software engineers develop GUIs following widespread architectural design patterns, such as MVC~\cite{Krasner88} or MVP~\cite{potel1996} (\emph{Model}-\emph{View}-\emph{Controller}/\emph{Presenter}), that consider GUIs as first-class concerns (\eg the \emph{View} in these two patterns).
These patterns clarify the implementations of GUIs by clearly separating concerns, thus minimizing the "spaghetti of call-backs"~\cite{Myers91}. 
These implementations rely on event-driven programming where events are treated by \emph{controllers} (resp. presenters\footnote{For simplicity, we use the term \emph{controller} to refer to any kind of component of MV* architectures that manages events triggered by GUIs, such as \emph{Presenter} (MVP), or \emph{ViewModel} (MVVM~\cite{smith09}).}), as depicted by \Cref{lst.introEX}.
In this code example, the \emph{AController} controller manages three widgets, \emph{b1}, \emph{b2}, and \emph{m3} (\Crefrange{codeIntro1}{codeIntro11}).
To handle events that these widgets trigger in response to users' interactions, the GUI listener $ActionListener$ is implemented in the controller (\Crefrange{codeIntro3}{codeIntro4}).
One major job of GUI listeners is the production of GUI commands, \ie a set of statements executed in reaction of a GUI event produced by a widget (\Cref{codeIntro5,codeIntro6,codeIntro10}).
Like any code artifact, GUI controllers must be tested, maintained and are prone to evolution and errors.
In particular, software developers are free to develop GUI listeners that can produce a single or multiple GUI commands.
In this work, we investigate the effects of such development practices on the code quality of the GUI listeners.

\begin{lstlisting}[xleftmargin=5.0ex,language=MyJava, label=lst.introEX, caption={Code example of a GUI controller}]
class AController implements ActionListener {
  JButton b1;|\label{codeIntro1}|
  JButton b2;|\label{codeIntro2}|
  JMenuItem m3;|\label{codeIntro11}|

  %%@Override%% public void actionPerformed(ActionEvent e){|\label{codeIntro3}|
     Object src = e.getSource();
     if(src==b1){|\label{codeIntro7}|
        // Command 1|\label{codeIntro5}|
     }else if(src==b2)|\label{codeIntro8}|
        // Command 2|\label{codeIntro6}|
     }else if(src instanceof AbstractButton && 
       ((AbstractButton)src).getActionCommand().equals(|\label{codeIntro9}|
         m3.getActionCommand()))
        // Command 3|\label{codeIntro10}|
     }
}}|\label{codeIntro4}|
\end{lstlisting}

\vspace*{-0.2cm}
\change{In many cases GUI code is intertwined with the rest of the code.
We thus propose a static code analysis required for detecting the GUI commands that a GUI listener can produce.
Using this code analysis, we then conduct a large empirical study on Java Swing open-source GUIs.
We focus on the Java Swing toolkit because of its popularity and the large quantity of Java Swing legacy code. 
We empirically study to what extent the number of GUI commands that a GUI listener can produce has an impact on the change- or fault-proneness of the GUI listener code,
considered in the literature as negative impacts of a design smell on the code~\cite{palomba14,Khomh2012,lozano2007assessing,rapu2004using}.
Based on the results of this experiment, we define a GUI design smell we call \bl, \ie  a GUI listener that can produce more than two GUI commands.
For example with \Cref{lst.introEX}, the GUI listener implemented in $AController$ manages events produced by three widgets, \emph{b1}, \emph{b2}, and \emph{m3} (\Cref{codeIntro7,codeIntro8,codeIntro9}), that produce one GUI command each.
21\% of the analyzed GUI controllers are \bls.
}

\change{We provide an open-source tool, \tool\footnote{\label{foot.webpage}\scriptsize\url{https://github.com/diverse-project/InspectorGuidget}}, that automatically detect \bls in Java Swing GUIs.
To evaluate  the ability of \tool at detecting \bls, we considered six representative Java software systems. 
We manually retrieved all instances of \bl in each application, to build a ground truth for our experiments: we found 37 \bls. 
\tool  detected 36 \bls out of 37. 
The experiments show that our algorithm has a precision of \SI{97.37}{\percent} and recall of \SI{97.59}{\percent} to detect \bls.
Our contributions are:}
\begin{itemize}\itemsep-0.03cm
  \item an empirical study on 13 Java Swing open-source software systems. 
  This study investigates the current coding practices of GUI controllers.
  The main result of this study is the identification of a GUI design smell we called \bl.
  \item a precise characterization of the \bl.
We also discuss the different coding practices of GUI listeners we observed in listeners having less than three commands.
  \item an open-source tool, \tool, that embeds a static code analysis to automatically detect the presence of \bls in Swing GUIs.
We evaluated the ability of \tool at detecting \bls.
\end{itemize}

The paper is organized as follows. 
\Cref{sec.study} describes an empirical study that investigates coding practices of GUI controllers.
Based on this study, \Cref{sec.goodBad} describes an original GUI design smell we called \bl.
Following, \Cref{sec.identify} introduces an algorithm to detect \bls, evaluated in \Cref{sec.eval}.
The paper ends with related work (\Cref{sec.related}) and a research agenda (\Cref{sec.conclu}).

\section{An empirical study on GUI listeners}\label{sec.study}

All the material of the experiments is freely available on the companion web page\footnoteref{foot.webpage}.

\subsection{Independent Variables}

GUI listeners are core code artifacts in software systems.
They receive and treat GUI events produced by users while interacting with GUIs.
In reaction of such events, GUI listeners produce GUI commands that can be defined as follows:

\begin{definition}[GUI Command]
A GUI command~\cite{GAM95,BEA00b}, aka. \emph{action}~\cite{BLO10,BLO11}, is a set of statements executed in reaction to a user interaction, captured by an input event, performed on a GUI. 
GUI commands may be supplemented with: 
a pre-condition checking whether the command fulfills the prerequisites to be executed;
undo and redo functions for, respectively, canceling and re-executing the command.
\end{definition}

\noindent\textbf{Number of GUI Commands} (\textbf{CMD}).
This variable measures the number of GUI commands a GUI listener can produce.
To measure this variable, we develop a dedicated static code analysis (see \Cref{sec.identify}).

GUI listeners are thus in charge of the relations between a software system and its GUI.
As any code artifact, GUI code has to be tested and analyzed to provide users with high quality (from a software engineering point of view) GUIs.
In this work, we specifically focus on a coding practice that affect the code quality of the GUI listeners:
we want to \emph{state whether the number of GUI commands that GUI listeners can produce has an effect on the code quality of these listeners}.
Indeed, software developers are free to develop GUI listeners that can produce a single or multiple GUI commands since no coding practices or GUI toolkits provide coding recommendations.
To do so, we study to what extent the number of GUI commands that a GUI listener can produce has an impact on the change- and fault-proneness of the GUI listener code.
Such a correlation has been already studied to evaluate the impact of several antipatterns on the code quality~\cite{Khomh2012}.

We formulate the research questions of this study as follows:
\begin{itemize}\itemsep-0.1cm
  \item[\textbf{RQ1}] To what extent the number of GUI commands per GUI listeners has an impact on fault-proneness of the GUI listener code?
  \item[\textbf{RQ2}] To what extent the number of GUI commands per GUI listeners has an impact on change-proneness of the GUI listener code?
  \item[\textbf{RQ3}] Does a threshold value, \ie a specific number of GUI commands per GUI listener, that can characterize a GUI design smell exist?
\end{itemize}

\subsection{Dependent Variables}

To answer the previously introduced research questions, we measure the following dependent variables.

\noindent\textbf{Average Commits} (\textbf{COMMIT}).
This variable measures the average number of commits per line of code (LoC) of GUI listeners.
This variable will permit to evaluate the change-proneness of GUI listeners.
The measure of this variable implies that the objects of this study, \ie software systems that will be analyzed, must have a large and accessible change history.
To measure this variable, we automatically count the number of the commits that concern each GUI listener.

\noindent\textbf{Average fault Fixes} (\textbf{FIX}). 
This variable measures the average number of fault fixes per LoC of GUI listeners.
This variable will permit to evaluate the fault-proneness of GUI listeners.
The measure of this variable implies that the objects of this study must use a large and accessible issue-tracking system.
To measure this variable, we manually analyze the log of the commits that concern each GUI listener.
We count the commits which log refers to a fault fix, \ie logs that point to a bug report of an issue-tracking system (using a bug ID or a URL) or that contain the term "\emph{fix}" (or a synonymous).

Both COMMIT and FIX rely on the ability to get the commits that concern a given GUI listener.
For each software system, we use all the commits of their history.
To identify the start and end lines of GUI listeners, we developed a static code analysis.
This code analysis uses the definition of the GUI listener methods provided by GUI toolkits (\eg \emph{void actionPerformed(ActionEvent)}) to locate these methods in the code. 
Moreover, commits may change the position of GUI listeners in the code (by adding or removing LoCs).
To get the exact position of a GUI listener while studying its change history, we use the Git tool \emph{git-log}\footnote{\scriptsize\url{https://git-scm.com/docs/git-log}}.
The \emph{git-log} tool has options that permit to: 
follow the file to log across file renames (option \emph{-M});
trace the evolution of a given line range across commits (option \emph{-L}).
We then manually check the logs for errors.

\subsection{Objects}

The objects of this study are a set of large open-source software systems.
The dependent variables, previously introduced, impose several constraints on the selection of these software systems.
They must use an issue-tracking system and the Git version control system.
Their change history must be large (\ie must have numerous commits) to let the analysis of the commits relevant.
In this work, we focus on Java Swing GUIs because of the popularity and the large quantity of Java Swing legacy code.
We thus selected from the Github platform\footnote{\scriptsize\url{https://github.com/}} \num{13} large Java Swing software systems.
The average number of commits of these software systems is approximately \num{2500} commits.
The total size of Java code is \num{1414}k Java LoCs.
Their average size is approximately \num{109}k Java LoCs.

\subsection{Results}

We can first highlight that the total number of GUI listeners producing at least one GUI command identified by our tool is \num{858}, \ie an average of \num{66} GUI listeners per software system.
This approximately corresponds to \num{20}~kLoCs, \ie around \SI{1.33}{\percent} of their Java code.

\begin{figure}[h]
	\centering
		\includegraphics[width=0.9\columnwidth]{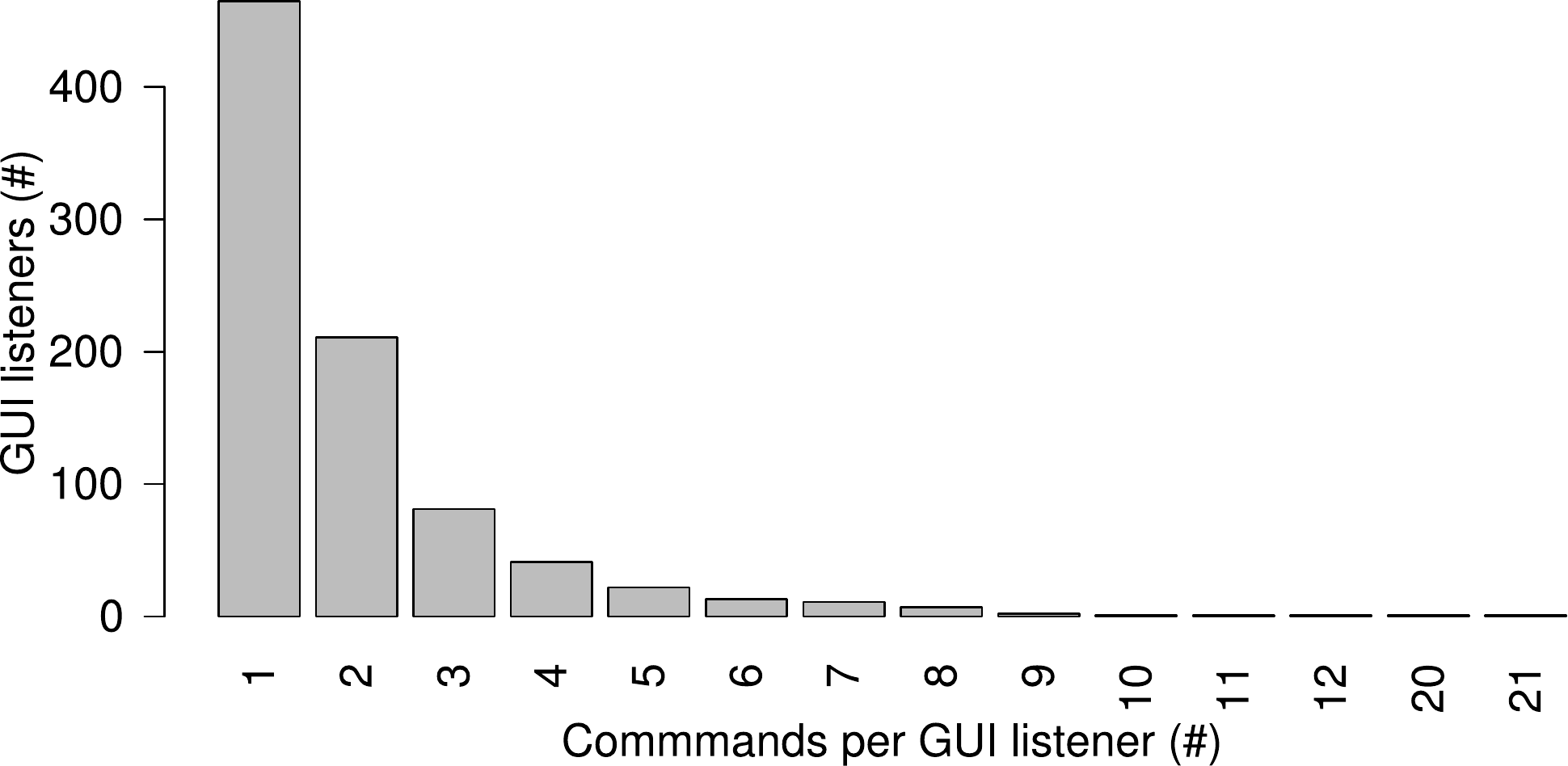}
		\caption{Distribution of the listeners according to their number of GUI commands}\label{fig.nbCmds}
\end{figure}

\Cref{fig.nbCmds} shows the distribution of the listeners according to their number of GUI commands.
Most of the listeners (\num{465}) can produce one command (we will call 1-command listener).
\num{211} listeners can produce two commands.
\num{81} listeners can produce three commands.
\num{101} listeners can produce at least four commands.
To obtain representative data results, we will consider in the following analyses four categories of listeners: 
\emph{one-command listener}, \emph{two-command listener}, \emph{three-command listener}, and \emph{four+-command listener}.

Besides, a first analysis of the data exhibits many outliers, in particular for the one-command listeners.
To understand the presence of these outliers, we manually scrutiny some of them and their change history.
We observe that some of these outliers are GUI listeners which size has been reduced over the commits.
For instance, we identified outliers that contained multiple GUI commands before commits that reduced them as one- or two-command listeners.
Such listeners distort the analysis of the results by considering listeners that have been large, as one- or two-command listeners.
We thus removed those outliers from the data set, since outliers removal, when justified, may bring benefits to the data analysis~\cite{Osborne2004}.
We compute the box plot statistics to identify and then remove the outliers.

\begin{figure}[h]
	\centering
		\includegraphics[width=0.97\columnwidth]{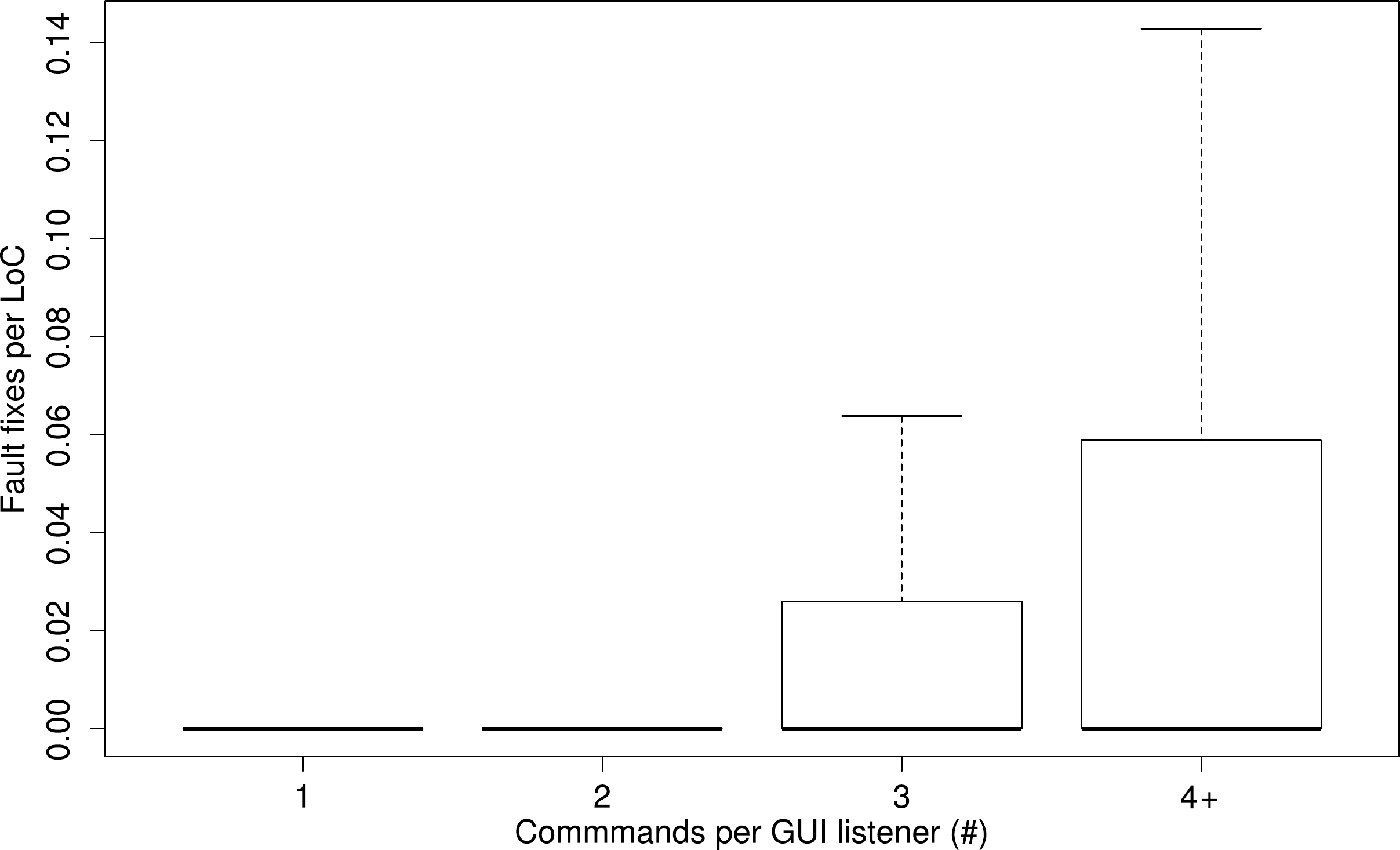}
		\caption{Number of bug fixes per LoC of GUI listeners}\label{fig.bugs}
\end{figure}

\Cref{fig.bugs} depicts the number of fault fixes per LoC (\ie \emph{FIX}) of the analyzed GUI listeners.
We observe an increase of the fault fixes per LoC when $CMD\ge 3$.
These results are detailed in \Cref{tab.results}.
The mean value of \emph{FIX} constantly increases over \emph{CMD}.
Because these data follow a monotonic relationship, we use the Spearman's rank-order correlation coefficient to assess the correlation between the number of fault fixes per LoC and the number of GUI commands in GUI listeners~\cite{Osborne2004}.
We also use a \SI{95}{\percent} confidence level (\ie $p$-value$<$0.05).
\change{This test exhibits a low correlation (\num{0.4438}) statistically significant with a $p$-value of \num{2.2e-16}.

\begin{table}[ht]\scriptsize
  \centering\setlength{\tabcolsep}{1.3pt}
  \caption{Mean, correlation, and significance of the results}\label{tab.results}
    \begin{tabular}{cSSSScc}
    \toprule
\textbf{Dependent} & \textbf{Mean} & \textbf{Mean} & \textbf{Mean} & \textbf{Mean} & \textbf{Correlation} & \textbf{Significance}\\
\textbf{variables} & \textbf{CMD=1} & \textbf{CMD=2} & \textbf{CMD=3} & \textbf{CMD>3} & & \textbf{$p$-value}  \\
 \cmidrule(lr){1-1} \cmidrule(lr){2-2} \cmidrule(lr){3-3} \cmidrule(lr){4-4} \cmidrule(lr){5-5} \cmidrule(lr){6-6} \cmidrule(lr){7-7}
$FIX$    & 0       & 0.0123 & 0.0190 & 0.0282 & \cellcolor{green} 0.4438  & \cellcolor{green} <0.001\\\midrule
$COMMIT$ & 0.0750  & 0.0767 & 0.0849 & 0.0576 & 0.0570  & 0.111\\\bottomrule
    \end{tabular}
\end{table}

Regarding \textbf{RQ1}, on the basis of these results we can conclude that \emph{the number of GUI commands per GUI listeners does not have a strong negative impact on fault-proneness of the GUI listener code.}
This result is surprising regarding the global increase that can be observed in \Cref{fig.bugs}.
One possible explanation is that the mean of the number of bugs per LoC slowly increases over the number of commands as shown in the first row of \Cref{tab.results}.
On the contrary, the range of the box plots of \Cref{fig.bugs} strongly increases with 3-command listeners.
This means that the 3+-command data sets are more variable than for the 1- and 2-command data sets.
}

\begin{figure}[h]
	\centering
		\includegraphics[width=0.99\columnwidth]{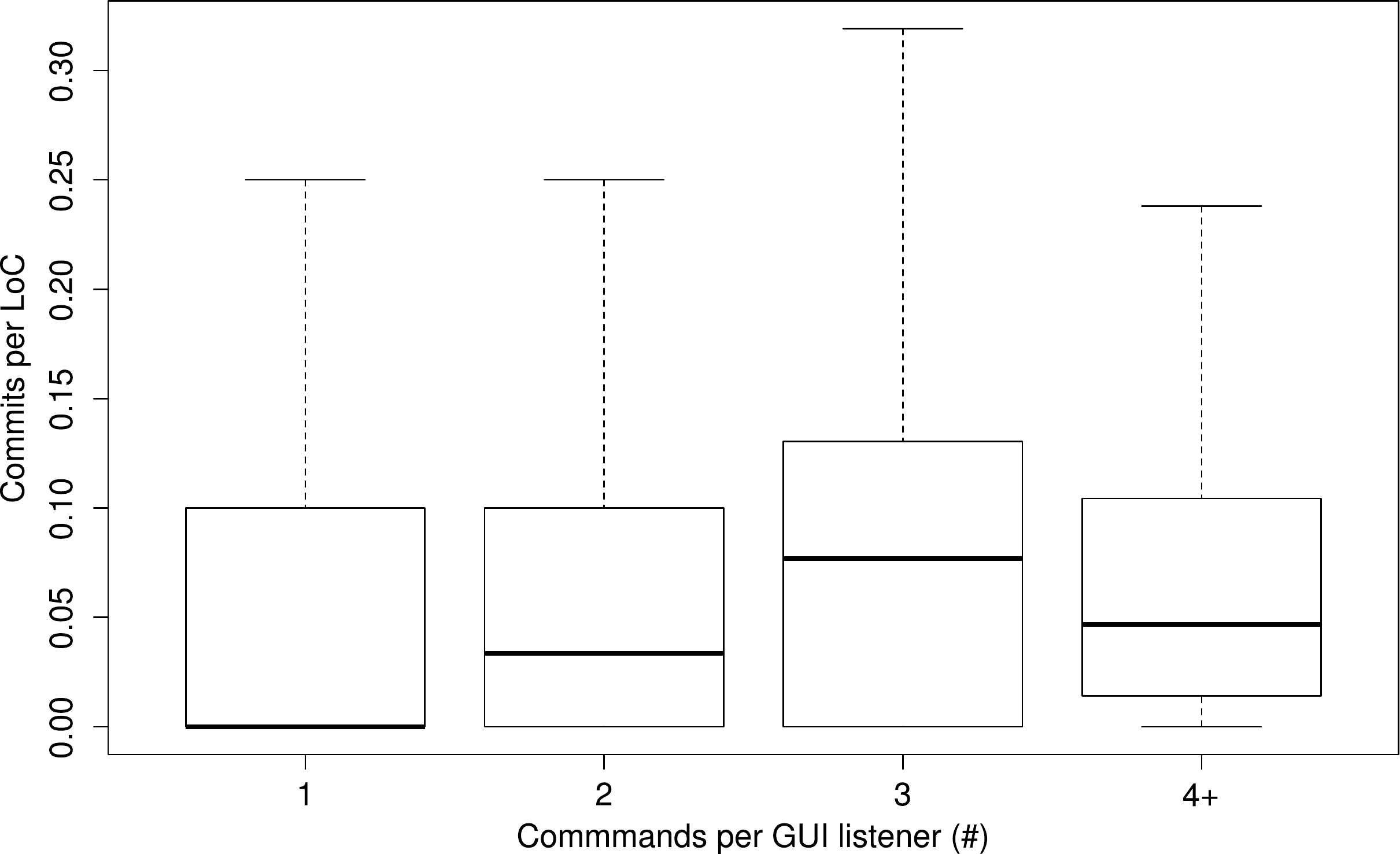}
		\caption{Number of commits per LoC of GUI listeners}\label{fig.commits}
\end{figure}

\Cref{fig.commits} depicts the number of commits per LoC (\ie \emph{COMMIT}) of the analyzed GUI listeners.
These results are also detailed in \Cref{tab.results}.
We observe that COMMIT does not constantly increases over CMD.
This observation is assessed by the absence of correlation between these two variables (\num{0.0570}), even if this result is not statistically significant with a $p$-value of \num{0.111}.
We can, however, observe in \Cref{fig.commits} an increase of COMMIT for the three-command listeners.

Regarding \textbf{RQ2}, on the basis of these results we can conclude that \emph{there is no evidence of a relationship between the number of GUI commands per GUI listeners and the change-proneness of the GUI listener code.}

\change{Regarding \textbf{RQ3}, we observe a significant increase of the fault fixes per LoC for 3+-command listeners.
We observe a mean of \num{0.004} bugs per LoC for 1- and 2-command listeners, against a mean of \num{0.024} bugs per LoC for 3+-command listeners, as highlighted by \Cref{fig.bugs}.
We apply the independent samples Mann-Whitney test to compare 1- and 2-command listeners against 3+-command listeners and we obtain a $p$-value of \num{2.2e-16} (\ie $p$-value$<$0.05).}
We observe similar, but not significant, increase on the commits per LoC for the three-command listeners.
We thus state that a threshold value, \ie a specific number of GUI commands per GUI listener, that characterizes a GUI design smell exists.
On the basis of the results, \emph{we define this threshold to three GUI commands per GUI listener}.
\change{Of course, this threshold value is an indication and as any design smell it may vary depending on the context.
Indeed, as noticed in several studies, threshold values of design smells must be customizable to let system experts the possibility to adjust them~\cite{Johnson2013,palomba2014}.
Using the threshold value of 3, the concerned GUI listeners represent 21\% of the analyzed GUI listener and \SI{0.54}{\percent} of the Java code of the analyzed software systems.}
Besides, the average size of the 3+-command listeners is \num{42} LoCs, \ie less than the \emph{long method} design smell defined between \num{100} and \num{150} LoCs in mainstream code analysis tools~\cite{Arcelli12}.

To conclude on this empirical study we highlight the main findings.
\change{The relation between the number of bug fixes over the number of GUI commands is too low to draw conclusions.
Future works will include a larger empirical study to investigate more in depth this relation.
However, a significant increase of the fault fixes per LoC for 3+-command listeners is observed.
We thus set to three the number of GUI commands beyond which a GUI listener is considered as badly designed.
This threshold value is an indication and as any design smell it may be defined by system experts according to the context.}
We show that \SI{0.54}{\percent} of the Java code of the analyzed software systems is affected by this new GUI design smell that concerns 21\% of the analyzed GUI listeners.
The threats to validity of this empirical study are discussed in \Cref{sub.threats}.



\section{Blob Listener: Definition \& Illustration}\label{sec.goodBad}

This section introduces the GUI design smell, we call \bl, identified in the previous section, and illustrates it through real examples.


\subsection{Blob Listener}\label{sub.blob}

We define the \bl as follows:

\begin{definition}[Blob Listener]\label{def.blob}
A \bl is a GUI listener that can produce more than two GUI commands.
\bls can produce several commands because of the multiple widgets they have to manage.
In such a case, \bls' methods (such as \emph{actionPerformed}) may be composed of a succession of conditional statements that:
1) check whether the widget that produced the event to treat is the good one, \ie the widget that responds a user interaction;
2) execute the command when the widget is identified.
\end{definition}

We identified three variants of the \bl.
The variations reside in the way of identifying the widget that produced the event.
These three variants are described and illustrated as follows.

\textbf{Comparing a property of the widget.}
\Cref{lst.fitsAllSwing} is an example of the first variant of \bl: 
the widgets that produced the event (\cref{code.3v,code.6v,code.8v}) are identified with a \emph{String} associated to the widget and returned by  \emph{getActionCommand} (\cref{code.2v}).
\change{Each of the three \emph{if} blocks forms a GUI \emph{command} to execute in response of on the triggered widget (\cref{code.4v,code.5v,code.7v,code.9v}).}

\begin{lstlisting}[xleftmargin=5.0ex,language=MyJava, caption={Widget identification using widget's properties in Swing}, label=lst.fitsAllSwing,escapechar=~]
public class MenuListener 
			   implements ActionListener, CaretListener {
 protected boolean selectedText;

 %%@Override%% public void actionPerformed(ActionEvent e) {
  Object src = e.getSource();
  if(src instanceof JMenuItem||src instanceof JButton){~\label{code.1v}~ 
		 String cmd = e.getActionCommand(); ~\label{code.2v}~ 
		 if(cmd.equals("Copy")){~\label{code.3v}~//Command #1 
			 if(selectedText)~\label{code.4v}~
				 output.copy();~\label{code.5v}~
		 }else if(cmd.equals("Cut")){~\label{code.6v}~//Command #2
			  output.cut();~\label{code.7v}~
		 }else if(cmd.equals("Paste")){~\label{code.8v}~//Command #3
			 output.paste();~\label{code.9v}~
		 }
		 // etc.
		}~\label{code.15v}~
  }
  %%@Override%% public void caretUpdate(CaretEvent e){
   	selectedText = e.getDot() != e.getMark();
   	updateStateOfMenus(selectedText);	
}}
\end{lstlisting}

In Java Swing, the properties used to identify widgets are mainly the \emph{name} or the \emph{action command} of these widgets.
The action command is a string used to identify the kind of commands the widget will trigger.
\Cref{lst.init}, related to \Cref{lst.fitsAllSwing}, shows how an action command (\cref{code.1b,code.3b}) and a listener (\cref{code.2b,code.4b}) can be associated to a widget in Java Swing during the creation of the user interface.

\begin{lstlisting}[xleftmargin=5.0ex,language=MyJava, caption={Initialization of Swing widgets to be controlled by the same listener}, label=lst.init]
menuItem = new JMenuItem();
menuItem.setActionCommand("Copy");|\label{code.1b}|
menuItem.addActionListener(listener);|\label{code.2b}|

button = new JButton();
button.setActionCommand("Cut");|\label{code.3b}|
button.addActionListener(listener);|\label{code.4b}|
//...
\end{lstlisting}

\textbf{Checking the type of the widget.}
The second variant of \bl consists of checking the \emph{type} of the widget that produced the event.
\Cref{lst.instanceof} depicts such a practice where the type of the widget is tested using the operator \emph{instanceof} (\Cref{code.1c,code.2c,code.3c,code.4c}).
One may note that such \emph{if} statements may have nested \emph{if} statements to test properties of the widget as explained in the previous point.
\begin{lstlisting}[xleftmargin=5.0ex,language=MyJava, caption={Widget identification using the operator \emph{instanceof}},label=lst.instanceof]
public void actionPerformed(ActionEvent evt) {
   Object target = evt.getSource();
   if (target instanceof JButton) {|\label{code.1c}|
      //...
   } else if (target instanceof JTextField) {|\label{code.2c}|
      //...
   } else if (target instanceof JCheckBox) {|\label{code.3c}|
      //...
   } else if (target instanceof JComboBox) {|\label{code.4c}|
      //...
}}
\end{lstlisting}

\textbf{Comparing widget references.}
The last variant of \bl consists of comparing widget references to identify those at the origin of the event.
\Cref{lst.fitsAllGWT} illustrates this variant where \emph{getSource} returns the source widget of the event that is compared to widget references contained by the listener (\eg \cref{code.1d,code.2d,code.3d}).

\begin{lstlisting}[xleftmargin=5.0ex,language=MyJava, caption={Comparing widget references}, label=lst.fitsAllGWT]
public void actionPerformed(ActionEvent event) {
   if(event.getSource() == view.moveDown) {|\label{code.1d}|
      //...
   } else if(event.getSource() == view.moveLeft) {|\label{code.2d}|
      //...
   } else if(event.getSource() == view.moveRight) {|\label{code.3d}|
      //...
   } else if(event.getSource() == view.moveUp) {
      //...
   } else if(event.getSource() == view.zoomIn) {
      //...
   } else if(event.getSource() == view.zoomOut) {
      //...
   }}
\end{lstlisting}

In these three variants, multiple \emph{if} statements are successively defined.
Such successions are required when one single GUI listener gathers events produced by several widgets.
In this case, the listener needs to identify the widget that produced the event to process.

The three variants of the \bl design smell also appear in others Java GUI toolkits, namely SWT, GWT, and JavaFX.
Examples for these toolkits are available on the companion webpage of this paper\footnoteref{foot.webpage}.

\section{Automatic Detection of GUI Commands\\and Blob Listeners}\label{sec.identify}

\subsection{Approach Overview}

\begin{figure}[h]
	\centering
 		\includegraphics[width=0.99\columnwidth]{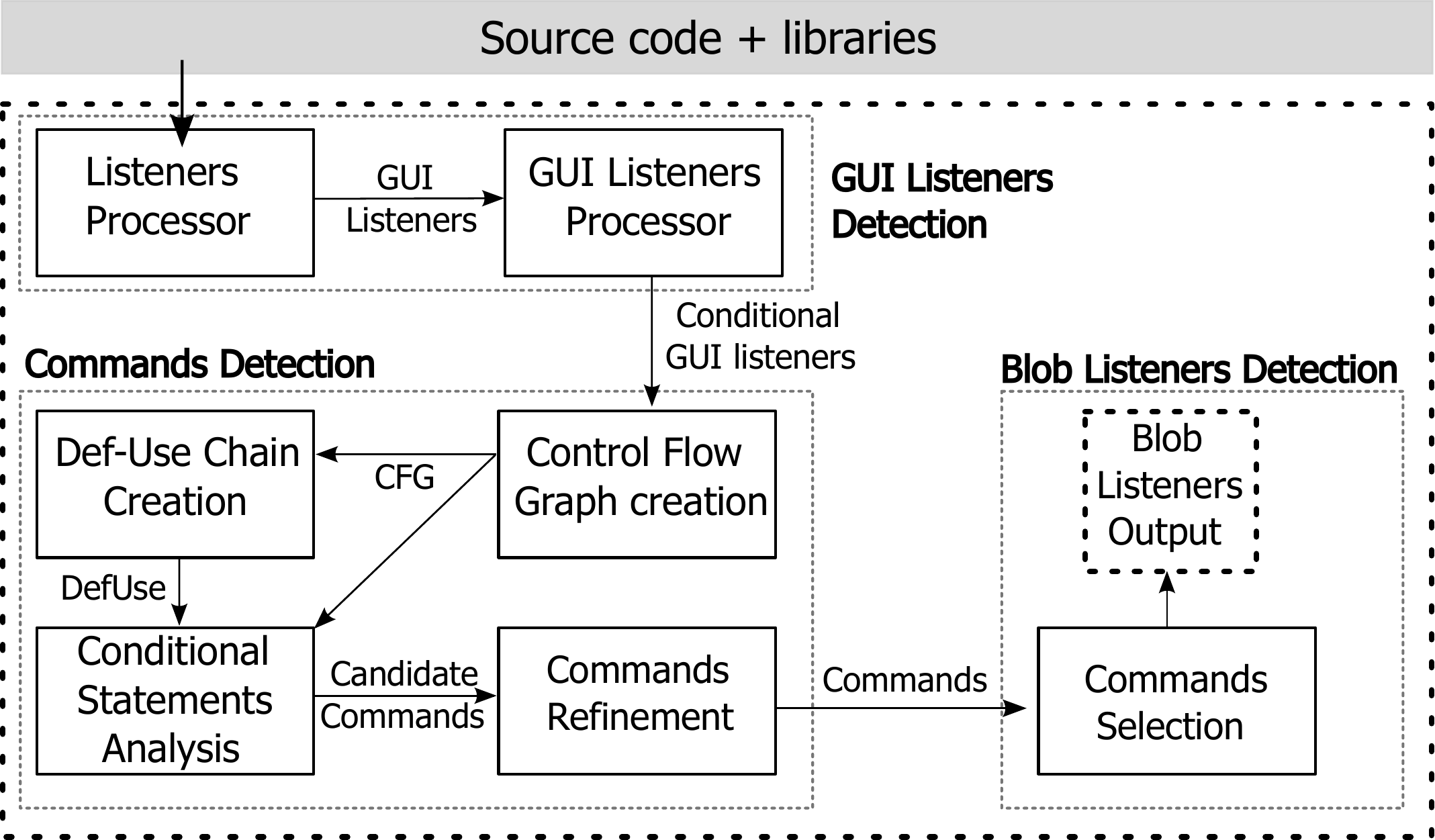}
		\caption{The proposed process for automatically detecting \bls}\label{fig.process}
\end{figure}

\Cref{fig.process} describes the process we propose to automatically detect \bls. 
The detection process includes three main steps.
First, GUI listeners that contain conditional blocks (conditional GUI listeners) are automatically detected in the source code through a static analysis (\Cref{sec.detectCondList}).
Then, the GUI commands, produced while interacting with widgets, that compose conditional GUI listeners are automatically detected using a second static analysis (\Cref{sec.extractCmd}).
This second static analysis permits to spot the GUI listeners that are \bls, \ie those having more than two commands.
\tool uses \emph{Spoon}, a library for transforming and analyzing Java source code~\cite{spoon}, to support the static analyses.

\subsection{Detecting Conditional GUI Listeners}\label{sec.detectCondList}

We define a conditional GUI listener as follows:
\begin{definition}[Conditional GUI listener]
A conditional GUI listener is a listener composed of conditional blocks used to identify the widget that produced an event to process.
Such conditional blocks may encapsulate a command to execute in reaction to the event.
\end{definition}

For instance, five nested conditional blocks (\Cref{code.1v,code.3v,code.4v,code.6v,code.8v}) compose the listener method \emph{actionPerformed} in \Cref{lst.fitsAllSwing} (\Cref{sec.goodBad}).
The first conditional block checks the type of the widget that produced the event (\Cref{code.1v}).
This block contains three other conditional blocks that identify the widget using its action command (\Cref{code.3v,code.6v,code.8v}).
Each of these three blocks encapsulates one command to execute in reaction of the event. 

\Cref{algo.detectList} details the detection of conditional GUI listeners.
The inputs are all the classes of an application and the  list of  classes of a GUI toolkit. 
First,  the source code classes are processed to identify the GUI controllers.
When a class implements a GUI listener (\Cref{code.3con}), all the implemented listener methods are retrieved (\Cref{code.4con}).
For example, a class that implements the \emph{MouseMotionListener} interface must implement the listener methods \emph{mouseDragged} and \emph{mouseMoved}.
Next, each GUI listener is analyzed to identify those having at least one conditional statement (\Cref{code.9con,code.10con}). 
All listeners with those statements are considered as conditional GUI listeners (\Cref{code.11con}).

\begin{algorithm} 
\caption{Conditional GUI Listeners Detection}\label{algo.detectList}  
\begin{algorithmic}[1]\small
\Require $classes$, the source classes of the software system
\Require $tkClasses$, the classes of the GUI toolkit
\Ensure $listeners$, the detected conditional GUI listeners
\State $GUIListeners \gets \varnothing$
\State $listeners \gets \varnothing$
\ForAll{$c \in classes$}\label{code.1con}
	\ForAll{$tkc \in tkClasses$}\label{code.2con}
		\If{$c$.isSubtypeOf($tkc$)}\label{code.3con}
			\State $GUIListeners \gets GUIListeners \cup getMethods(c, tkc)$\label{code.4con}
		\EndIf
  	\EndFor\label{code.6con}
\EndFor
\ForAll{$listener \in GUIListeners$}\label{code.8con}
		\State $statements \gets getStatements(listener)$\label{code.9con}
			\If{hasConditional($statements$)}\label{code.10con}
				\State $listeners \gets listeners \cup \{listener\}$\label{code.11con}
			\EndIf	
\EndFor
\end{algorithmic}
\end{algorithm}

\subsection{Detecting Commands in Conditional GUI Listeners}\label{sec.extractCmd}

\Cref{algo.command} details the detection of GUI commands.
The input is a set of GUI conditional listeners. 
The statements of conditional GUI listeners are processed to detect commands.
First, we build the control-flow graph (CFG) of each listener (\Cref{code.3co}).
Second, we traverse the CFG to gather all the conditional statements that compose a given statement (\Cref{code.4co}).
Next, these conditional statements are analyzed to detect any reference to a GUI event or widget (\Cref{code.5co}).
Typical references we found are for instance:

\begin{lstlisting}[language=MyJava,numbers=none]
if(e.getSource() instanceof Component)...
if(e.getSource() == copy)...
if(e.getActionCommand().contains("copy"))...
\end{lstlisting}

where \emph{e} refers to a GUI event, \emph{Component} to a Swing class, and \emph{copy} to a Swing widget.
The algorithm recursively analyzes the variables and class attributes used in the conditional statements until a reference to a GUI object is found in the controller class.
For instance, the variable \emph{actionCmd} in the following code excerpt is also considered by the algorithm.

\begin{lstlisting}[language=MyJava,numbers=none]
String actionCmd = e.getSource().getActionCommand()
if("copy".equals(actionCmd)) ...
\end{lstlisting}

When a reference to a GUI object is found in a conditional statement, it is considered as a potential command (\Cref{code.6co}).
These potential commands are then more precisely analyzed to remove irrelevant ones (\Crefrange{code.59co}{code.76co}) as discussed below.

\begin{algorithm} 
\caption{Commands Detection}\label{algo.command}  
\begin{algorithmic}[1]\small
\Require $listeners$, the detected conditional GUI listeners
\Ensure $commands$, the commands detected in $listeners$
\State $commands \gets getProperCmds(getPotentialCmds(listeners))$
\State
\Function{getPotentialCmds}{$listeners$}\label{code.1co}	
   \State $cmds = \varnothing$
	\ForAll{$listener \in listeners$}\label{code.2co}		
		\State $cfg \gets getControlFlowGraph(listener)$\label{code.3co}
		\ForAll{$stmts \in cfg$}\label{code.4co}
			\State $conds \gets$ $getCondStatementsUseEventOrWidget$($stmts$)\label{code.5co}
			\State $cmds = cmds \cup \{Command(stmts, conds, listener)\}$\label{code.6co}
		\EndFor\label{code.7co}	
     	\EndFor
     	\State \Return $cmds$\label{code.8co}
\EndFunction
\State
\Function{getProperCmds}{$candidates$}\label{code.59co}
   \State $nestedCmds \gets \varnothing$
   \State $notCandidates \gets \varnothing$
	\ForAll{$cmd \in candidates$}\label{code.60co}	
		\State $nestedCmds \gets nestedCmds  \cup \left(cmd, getNestCmds(cmd)\right)$\label{code.61co}	
      \EndFor
	\ForAll{$(cmd, nested) \in nestedCmds, |nested|>0$}\label{code.63co}
		\If{$|nested| == 1$}\label{code.64co}
			\State $notCandidates \gets notCandidates \cup nested$\label{code.65co}
			\Else
				\State $notCandidates \gets notCandidates \cup \{cmd\}$\label{code.70co}
		\EndIf
     	\EndFor\label{code.73co}
	\State \Return $candidates \cap notCandidates$\label{code.74co}
\EndFunction\label{code.76co}
\end{algorithmic}
\end{algorithm}

A conditional block statement can be surrounded by other conditional blocks.
Potential commands detected in the function \emph{getPotentialCmds} can thus be nested within other commands.
We define such commands as \emph{nested commands}.
In such a case, the algorithm analyzes the nested conditional blocks to detect the most representative command.
We observed two cases:
\begin{inparaenum}\itemsep0cm
\item A potential command contains only a single potential command, recursively.
The following code excerpt depicts this case.
Two potential commands compose this code. 
Command~\#1 (\Crefrange{code.100}{code.104}) has a set of statements (\ie{} command~\#2) to be executed when the widget labeled \emph{"Copy"} is \emph{pressed}.
However, command~\#2 (\Crefrange{code.101}{code.103}) only checks whether there is a text typed into the widget \emph{"output"} to then allow the execution of  command \#1.
So, command~\#2 works as a precondition to command~\#1,  which is the command executed in reaction to that interaction.
In this case, only the first one will be considered as a GUI command.  
\begin{lstlisting}[xleftmargin=5.0ex,language=MyJava] 
if(cmd.equals("Copy")){ //Potential command #1|\label{code.100}|
 if(!output.getText().isEmpty()){//Potential command #2|\label{code.101}| 
  output.copy();|\label{code.102}|
 }|\label{code.103}|
}|\label{code.104}|
\end{lstlisting}
\item A potential command contains more than one potential command.
The following code excerpt depicts this case.
Four potential commands compose this code (\Cref{code.110,code.111,code.112,code.113}).
In this case, the potential commands that contain multiple commands are not considered.
In our example, the first potential command (\Cref{code.110}) is ignored.
One may note that this command checks the type of the widget, which is a variant of \bl (see \Cref{sub.blob}).
The three nested commands, however, are the real commands triggered on user interactions.
\begin{lstlisting}[xleftmargin=5.0ex,language=MyJava,escapechar=~]
if(src instanceof JMenuItem){ //Potential command #1~\label{code.110}~
	String cmd = e.getActionCommand();
	if(cmd.equals("Copy")){ //Potential command #2~\label{code.111}~
	}
	else if(cmd.equals("Cut")){ //Potential command #3~\label{code.112}~
	}
	else if(cmd.equals("Paste")){ //Potential command #4~\label{code.113}~
	}
}
\end{lstlisting}
\end{inparaenum}

These two cases are described in \Cref{algo.command} (\Crefrange{code.63co}{code.70co}).
Given a potential command, all its nested potential commands are gathered (\Crefrange{code.60co}{code.61co}).
The function \emph{getNestCmds} analyzes the commands by comparing their code line positions, statements, \etc{} 
So, if one command $C$ contains other commands, they are marked as nested to $C$.
Then, for each potential command and its nested ones: if the number of nested commands equals 1, the single nested command is ignored (\Crefrange{code.64co}{code.65co}); 
if the number of nested commands is greater than 1, the root command is ignored (\Cref{code.70co}).
\change{Finally, GUI listeners that can produce more than two commands are marked as \bls.
\tool allows the setting of this threshold value to let system experts the possibility to adjust them, as suggested by several studies~\cite{Johnson2013,palomba2014}.}

\begin{table*}[h]\scriptsize
 \caption{Selected interactive systems and some of their characteristics}\label{tab.appSize} 
  \centering \setlength{\tabcolsep}{3pt}
    \begin{tabular}{llccl}
    \toprule
\textbf{Software}&   \textbf{Version} &\textbf{GUI}                                 &\textbf{Conditional}            &\textbf{Source Repository Link}\\
\textbf{Systems}  &                                &\textbf{Listeners (\# (LoC))}   &\textbf{GUI Listeners (\# (LoC))}  & \\
 	\cmidrule(lr){1-1}\cmidrule(lr){2-2}\cmidrule(lr){3-3}\cmidrule(lr){4-4}         \cmidrule(lr){5-5}
FastPhotoTagger&2.3                             &94~(555)                       &23 (408)                       &\url{http://sourceforge.net/projects/fastphototagger/}\\
GanttProject        &2.0.10                       &67~(432)                       &14 (282)                        &\url{https://code.google.com/p/ganttproject/}\\
JaxoDraw            &2.1                           &123~(1331)                      &50 (1128)                       &\url{http://jaxodraw.svn.sourceforge.net/svnroot/jaxodraw/trunk/jaxodraw}\\
Jmol                    &14.1.13                    &248~(1668)                      &53 (1204)                       &\url{http://svn.code.sf.net/p/jmol/code/trunk/Jmol}\\
TerpPaint             &3.4                           &272~(1089)                      &4 (548)                         &\url{http://sourceforge.net/projects/terppaint/files/terppaint/3.4/}\\
TripleA               &1.8.0.3                  &559~(6138)                    &174 (4321)                        &\url{https://svn.code.sf.net/p/triplea/code/trunk/triplea/}\\
	\bottomrule
    \end{tabular}
\end{table*}

\section{Evaluation}\label{sec.eval}

To evaluate the efficiency of our detection algorithm, we address the two following research questions:
\begin{itemize}\itemsep0cm
  \item[\textbf{RQ4}] To what extent is the detection algorithm able to detect GUI commands in GUI listeners correctly?
  \item[\textbf{RQ5}] To what extent is the detection algorithm able to detect \bls correctly?
\end{itemize}

The evaluation has been conducted using \tool, our implementation of the \bl detection algorithm.
\tool is an Eclipse plug-in that analyzes Java Swing software systems.
\tool leverages the Eclipse development environment to raise warnings in the Eclipse Java editor on detected Blob listeners and their GUI commands.
Initial tests have been conducted on software systems not reused in this evaluation.
\tool and all the material of the evaluation are freely available on the companion web page\footnoteref{foot.webpage}.

\subsection{Objects}

We conducted our evaluation by selecting six well-known or large open-source software systems based on the Java Swing toolkit:
\emph{FastPhotoTagger}, \emph{GanttProject}, \emph{JAxoDraw}, \emph{Jmol}, \emph{TerpPaint}, and \emph{TripleA}.
We use other software systems than those used in our empirical study (\Cref{sec.study}) to diversify the data set used in this work and assess the validation of the detection algorithm on other systems.
Only \emph{GanttProject} is part of both experiments since it is traditionally used in experiments on design smells.
\Cref{tab.appSize} lists these systems and some of their characteristics such as their number of GUI listeners.

\subsection{Methodology}
The accuracy of the static analyses that compose the detection algorithm is measured by the \emph{recall} and \emph{precision} metrics~\cite{Frolin15}.
We ran \tool on each software system to detect GUI listeners, commands, and \bls.
We assume as a precondition that only GUI listeners are correctly identified by our tool.
Thus, to measure the precision and recall of our automated approach, we manually analyzed all the GUI listeners detected by \tool to:
\begin{itemize}\itemsep0cm
\item \emph{Check conditional GUI Listeners}. 
For each GUI listener, we manually checked whether it contains at least one conditional GUI statement. 
The goal is to answer RQ4 and RQ5 more precisely, by verifying whether all the conditional GUI listeners are statically analyzed to detect commands and \bls. 
\item \emph{Check commands}. We analyzed the conditional statements of GUI listeners to check whether they encompass commands.
Then, \emph{recall} measures the percentage of relevant commands that are detected (\Cref{eq.recallCmd}). 
\emph{Precision} measures the percentage of detected commands that are relevant (\Cref{eq.precCmd}).

\begin{equation}\textstyle\label{eq.recallCmd}\small
Recall_{cmd} (\%) = \frac{|\{ RelevantCmds\}  \cap \{ DetectedCmds\} |}{|\{ RelevantCmds\} |} \times 100   
\end{equation}

\begin{equation}\textstyle\label{eq.precCmd}\small
Precision_{cmd} (\%) = \frac{|\{ RelevantCmds\} \cap \{ DetectedCmds\}|}{|\{ DetectedCmds\}|} \times 100
\end{equation}

\emph{RelevantCmds} corresponds to  all the commands defined in GUI listeners, \ie the commands that should be detected by \tool.
\emph{Recall} and \emph{precision} are calculated over the number of false positives (FP) and false negatives (FN).
A command incorrectly detected by \tool while it is not a command, is classified as false positive.
A false negative is a command not detected by \tool.

\item \emph{Check \bls.}
To check whether a GUI listener is a \bl, we stated whether the commands it contains concern several widgets.
We use the same metrics of commands detection to measure the accuracy of \bls detection:

\begin{equation}\textstyle\label{eq.recallBlob}\small
Recall_{blob} (\%) = \frac{|\{ RelevantBlobs\} \cap \{ DetectedBlobs\}|}{|\{ RelevantBlobs\}|} \times 100
\end{equation}

\begin{equation}\textstyle\label{eq.precBlob}\small
Precision_{blob} (\%) = \frac{|\{ RelevantBlobs\} \cap \{ DetectedBlobs\}|}{|\{ DetectedBlobs\}|} \times 100
\end{equation}

Relevant \bls are all the GUI listeners that handle more than two commands (see \Cref{sec.identify}). 
Detecting \bls is therefore dependent on the commands detection accuracy.
\end{itemize}

\subsection{Results and Analysis}\label{sub.results}
\noindent\textbf{RQ4: Command Detection Accuracy.} 
\Cref{tab.evalCmd} shows the number of commands successfully detected per software system.
\emph{TripleA} has presented the highest number of GUI listeners (559), conditional GUI listeners (174), and commands (152).
One can notice that despite the low number of conditional GUI listeners that has \emph{TerpPaint} (4), this software system has 34 detected commands.
So, according to the sample we studied, the number of commands does not seem to be correlated to the number of conditional GUI listeners.

\begin{table}[H]\scriptsize
 \caption{Command Detection Results}\label{tab.evalCmd}
  \centering \setlength{\tabcolsep}{3.4pt}
    \begin{tabular}{lSSSSS} 
    \toprule
	\textbf{Software} & \textbf{Successfully} 			&\textbf{FN} 	&\textbf{FP} 	&$\mathbf{Recall_{cmd}}$& $\mathbf{Precision_{cmd}}$\\
		\textbf{System}	   & \textbf{Detected}&\textbf{(\#)}	&\textbf{(\#)} &\textbf{(\%)}			    &\textbf{(\%)}\\	
	                  	& \textbf{Commands (\#)}     & & &  & \\	
 	\cmidrule(lr){1-1} \cmidrule(lr){2-2} \cmidrule(lr){3-3} \cmidrule(lr){4-4}  \cmidrule(lr){5-5} \cmidrule(lr){6-6}
  	FastPhotoTagger&30&4&0&88.24&100.00\\
  	GanttProject&19&6&0&76.00&100.00\\
  	JaxoDraw&99&3&2&97.06&98.02\\
  	Jmol&103&18&2&85.12&98.10\\
  	TerpPaint&34&1&0&97.14&100.00\\
  	TripleA&152&44&0&77.55&100.00\\
	\cmidrule(lr){1-6}
	OVERALL&437&76&4&85.89&99.10\\
	\bottomrule
   \end{tabular}
\end{table}

\Cref{tab.evalCmd} also reports the number of FN and FP commands, and the values of the \emph{recall} and \emph{precision} metrics.
\emph{TripleA} and \emph{Jmol} revealed the highest number of FN, whereas \emph{TerpPaint} presented the lowest number of FN. 
The \emph{precision} of the command detection is \SI{99.10}{\percent}.
Most of the commands (437/441) detected by our algorithm are relevant.
We, however, noticed 76 relevant commands not detected leading to an average \emph{recall} of \SI{85.89}{\percent}.
Thus, our algorithm is less accurate in detecting all the commands than in detecting the relevant ones.
For example, \emph{TripleA} revealed 44 FN commands and no false positive result, leading to a recall of \SI{77.55}{\percent} and a precision of \SI{100}{\percent}.
The four FP commands has been observed in \emph{JAxoDraw} (2) and \emph{Jmol} (2), leading to a precision of \SI{98.02}{\percent} and \SI{98.10}{\percent} respectively.

\begin{figure}[h]
	\centering
		\includegraphics[width=0.99\columnwidth]{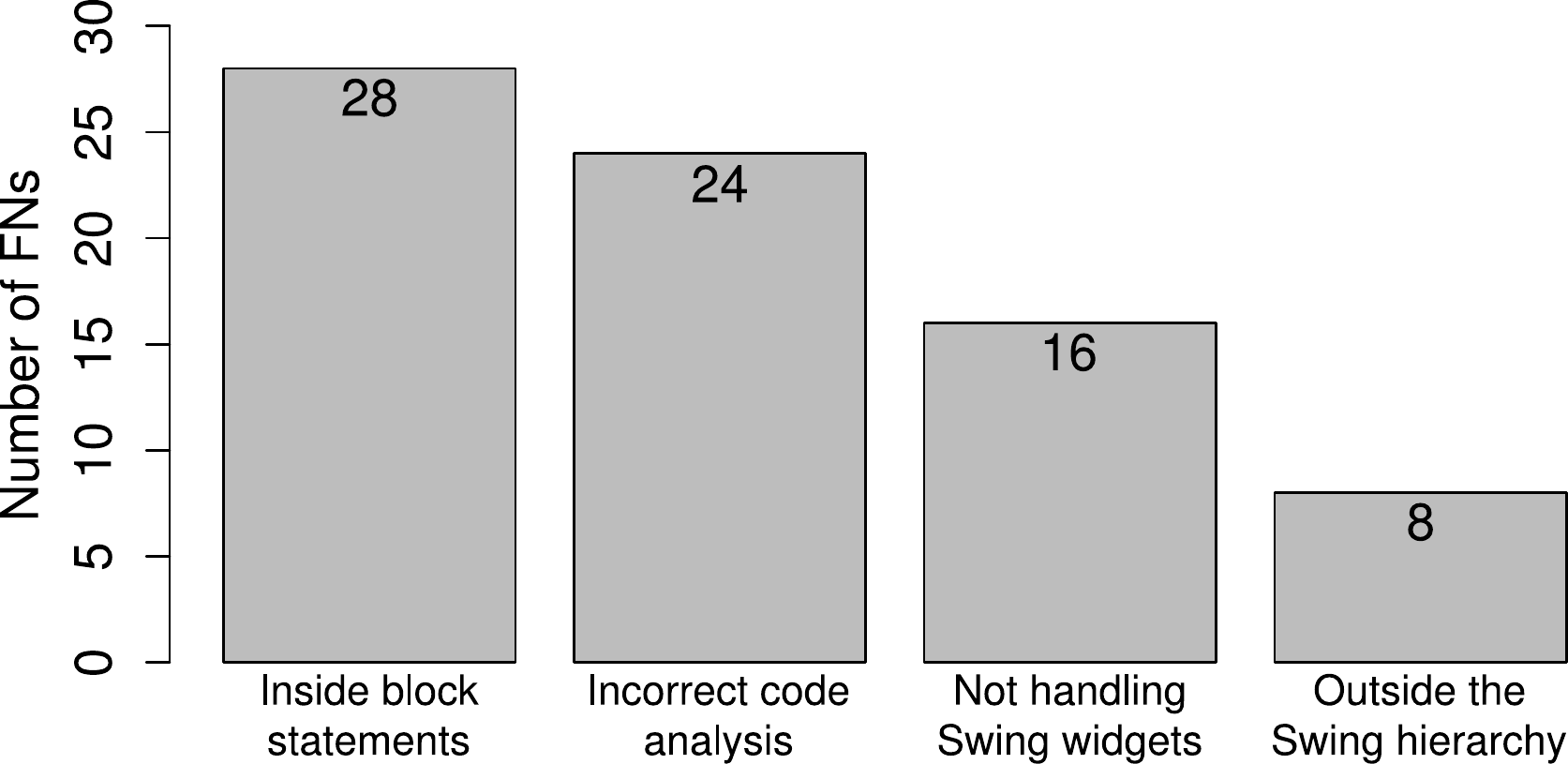}
		\caption{Distribution of the false negative commands}\label{fig.fnCmds}
\end{figure}

\Cref{fig.fnCmds} classifies the 76 FN commands according to the cause of their non-detection.
28 commands were not detected because of the use of widgets \emph{inside block statements} rather than inside the conditional statements.
For example, their conditional expressions refer to boolean or integer types rather than widget or event types.
16 other commands were not detected since they rely on \adhoc widgets or GUI listeners.
These widgets are developed for a specific purpose and rely on specific user interactions and complex data representation~\cite{LEL15}.
Thus, our approach cannot identify widgets that are not developed under Java Swing toolkit.
All the FN commands reported in this category concern \emph{TripleA} (14) and \emph{Jmol} (2) that use several \adhoc widgets.
Similarly, we found eight FN commands that use classes defined \emph{outside the Swing class hierarchy}.
A typical example is the use of widgets' models (\eg classes \emph{ButtonModel} or \emph{TableModel}) in GUI listeners.
Also, we identified 24 FN commands caused by an \emph{incorrect code analysis} (either bugs in \tool or in the \emph{Spoon} library).
This result was mainly affected by \emph{Jmol}, that has a listener with 14 commands not detected.

To conclude on RQ4, our approach is efficient for detecting GUI commands that compose GUI listener, even if some improvements are possible.

\noindent\textbf{RQ5: Blob Listeners Detection Accuracy.}
\Cref{tab.evalBlobs} gives an overview of the results of the \bls detection per software system.
The highest numbers of detected \bls concern \emph{TripleA} (11), \emph{Jmol} (11), and \emph{JAxoDraw} (7).
\change{Only one false positive and false negative have been identified against 37 \bls correctly detected.
The average recall is \SI{97.59}{\percent} and the average precision is \SI{97.37}{\percent}.
The average time spent to analyze the software systems is \SI{10810}{\milli\second}.
It includes the time that \emph{Spoon} takes to process all classes plus the time to detect GUI commands and \bls.
The worst-case is measured in \emph{TripleA}, \ie the largest system, with \SI{16732}{\milli\second}.
\emph{Spoon} takes a significant time to load the classes for large software systems (\eg \SI{12437}{\milli\second} out of \SI{16732}{\milli\second} in \emph{TripleA}).
Similarly to the command detection, we did not observe a correlation between the number of conditional GUI listeners, commands, and \bls.
So, regarding the recall and the precision, our approach is efficient for detecting \bls.}

\begin{table}[h]\scriptsize
 \caption{Blob Listener Detection Results}\label{tab.evalBlobs}
  \centering \setlength{\tabcolsep}{1.7pt}
    \begin{tabular}{lSSSSSr} 
    \toprule
	\textbf{Software} & \textbf{Successfully} 	  &\textbf{FN}  &\textbf{FP}    &$\mathbf{Recall_{blob}}$&$\mathbf{Precision_{blob}}$ &\textbf{Time} \\
  	\textbf{System}   & \textbf{Detected}	         &\textbf{(\#)}&\textbf{(\#)}  &\textbf{(\%)}		      &\textbf{(\%)}			         &\textbf{(ms)}\\
  	& \textbf{\bls (\#)} & & & & & \\	
 	\cmidrule(lr){1-1} \cmidrule(lr){2-2} \cmidrule(lr){3-3} \cmidrule(lr){4-4}  \cmidrule(lr){5-5} \cmidrule(lr){6-6}\cmidrule(lr){7-7}
  	FastPhotoTagger&3&0&0&100.00&100.00&3445\\
  	GanttProject&2&0&0&100.00&100.00&1910\\
  	JAxoDraw&7&0&1&100.00&87.50&13143\\
  	Jmol&11&1&0&91.67&100.00&16904\\
  	TerpPaint&3&0&0&100.00&100.00&12723\\
  	TripleA&11&0&0&100.00&100.00&16732\\
	\cmidrule(lr){1-7}
	OVERALL&37&1&1&97.59&97.37&10810\\
	\bottomrule
   \end{tabular}
\end{table}

\change{Regarding the single FN \bl, located in the \emph{Jmol} software system, this FN is due to an error in our implementation.
Because of a problem in the analysis of variables in the code, 14 GUI commands were not detected.}
\Cref{lst.codeJaxoDraw} gives an example of the FP \bl detected in \emph{JAxoDraw}.
It is composed of  three commands based on checking the \emph{states of widgets}.
For instance, the three commands rely on the selection of a list (Lines \ref{code.4t}, \ref{code.7t}, and \ref{code.10t}).

\begin{lstlisting}[xleftmargin=5.0ex,language=MyJava, caption={GUI code excerpt, from \emph{JAxoDraw}}, label=lst.codeJaxoDraw, escapechar=~]
public final void valueChanged(ListSelectionEvent e) {
	if (!e.getValueIsAdjusting()) {
		final int index = list.getSelectedIndex();
		if (index == -1) {~\label{code.4t}~//Command #1
			removeButton.setEnabled(false);
			packageName.setText("");
		} else if ((index == 0)||(index == 1){~\label{code.7t}~//Command #2
				 || (index == 2))
				removeButton.setEnabled(false);
				packageName.setText("");
		} else {~\label{code.10t}~//Command #3
			removeButton.setEnabled(true);
			String name = list.getSelectedValue().toString();
			packageName.setText(name);
		}
}}
\end{lstlisting}

\section{Discussion}\label{sub.disc}
In the next three subsections, we discuss the threats to validity of the experiments detailed in this paper, the scope of \tool, and alternative coding practices that can be used to limit \bls.

\subsection{Threats to validity} \label{sub.threats}

\textbf{External validity.} 
This threat concerns the possibility to generalize our findings.
We designed the experiments using multiple Java Swing open-source software systems to diversify the observations.
These unrelated software systems are developed by different persons and cover various user interactions.
Several selected software systems have been used in previous research work, \eg{} \emph{GanttProject}~\cite{Arcelli12, ADA11}, \emph{Jmol}~\cite{ADA11}, and \emph{TerpPaint}~\cite{COH12} that have been extensively used against GUI testing tools.
Our implementation and our empirical study (\Cref{sec.study}) focus on the Java Swing toolkit only.
We focus on the Java Swing toolkit because of its popularity and the large quantity of Java Swing legacy code.
We provide on the companion web page examples of \bls in other Java GUI toolkits, namely GWT, SWT, and JavaFX\footnoteref{foot.webpage}.

\textbf{Construct validity.}
This threat relates to the perceived overall validity of the experiments.
Regarding the empirical study (\Cref{sec.study}), we used \tool to find GUI commands in the code.
\tool might not have detected all the GUI commands.
We show in the validation of this tool (\Cref{sec.eval}) that its precision (\num{99.10}) and recall (\num{86.05}) limit this threat.
Regarding the validation of \tool, the detection of FNs and FPs have required a manual analysis of all the GUI listeners of the software systems.
To limit errors during this manual analysis, we added a debugging feature in \tool for highlighting GUI listeners in the code.
We used this feature to browse all the GUI listeners and identify their commands to state whether these listeners are \bls.
During our manual analysis, we did not notice any error in the GUI listener detection.
We also manually determined whether a listener is a \bl.
To reduce this threat, we carefully inspected each GUI command highlighted by our tool.

\subsection{Scope of the Approach}
Our approach has the following limitations.
First, \tool currently focuses on GUIs developed using the Java \emph{Swing} toolkit.
This is a design decision since we leverage Spoon, \ie{} a library to analyze \emph{Java} source code.
However, our solution is generic and can be used to support other GUI toolkits. 

Second, our solution is limited to analyze GUI listeners and associated class attributes. 
We identified several GUI listeners that dispatch the event processing to methods. 
Our implemented static analyses can be extended to traverse these methods to improve its performance.

Last, the criteria for the \bls detection should be augmented by inferring the related commands.
For example, when a GUI listener is a \bl candidate, our algorithm should analyze its commands by comparing their commonalities (\eg shared widgets and methods).
The goal is to detect commands that form in fact a single command.


\subsection{Alternative Practices}\label{sub.good}

We scrutinized GUI listeners that are not \bls to identify alternative practices that may limit \bls.
In most of the cases, these practices consist of producing one command \emph{per} listener by managing \emph{one} widget per listener.

\textbf{Listeners as anonymous classes.} \Cref{lst.good} is an example of this good practice.
A listener, defined as an anonymous class (\Crefrange{code.1e}{code.2e}), registers with one widget (\Cref{code.0e}).
The methods of this listener are then implemented to define the command to perform when an event occurs.
Because such listeners have to handle only one widget, \emph{if} statements used to identify the involved widget are not more used, simplifying the code.

\begin{lstlisting}[xleftmargin=5.0ex,language=MyJava, caption={Good practice for defining controllers: one widget per listener}, label=lst.good]
private void registerWidgetHandlers() {
 view.resetPageButton().addActionListener(|\label{code.0e}|
   new ActionListener() {|\label{code.1e}|
   %%@Override%% 
   public void actionPerformed(ActionEvent e){
     requestData(pageSize, null);
   }
 });|\label{code.2e}|
 view.previousPageButton().addActionListener(
   new ActionListener() {
     %%@Override%% 
     public void actionPerformed(ActionEvent e){
       if(hasPreviousBookmark())
         requestData(pageSize, getPreviousBookmark());
     }
   });//...
}
\end{lstlisting}

\vspace*{-0.2cm}
\textbf{Listeners as lambdas.} \Cref{lst.goodJava8} illustrates the same code than \Cref{lst.good} but using Lambdas supported since Java 8.
Lambdas simplify the implementation of anonymous class that have a single method to implement.

\begin{lstlisting}[xleftmargin=5.0ex,language=MyJava, caption={Same code than in Listing~\ref{lst.good} but using Java 8 Lambdas}, label=lst.goodJava8]
private void registerWidgetHandlers() {
 view.resetPageButton().addActionListener(
              e -> requestData(pageSize, null));|\label{code.0f}|
   
 view.previousPageButton().addActionListener(e -> {
  if (hasPreviousBookmark())
    requestData(pageSize, getPreviousBookmark());
 });
 
 //...
}
\end{lstlisting}

\vspace*{-0.2cm}
\textbf{Listeners as classes.} In some cases, listeners have to manage different intertwined methods.
This case notably appears when developers want to combine several listeners or methods of a single listener to develop a more complex user interaction.
For example, \Cref{lst.listClass} is a code excerpt that describes a mouse listener where different methods are managed:
\emph{mouseClicked} (\Cref{code.0g}), \emph{mouseReleased} (\Cref{code.1g}), and \emph{mouseEntered} (\Cref{code.2g}).
Data are shared among these methods (\emph{isDrag}, \Cref{code.3g,code.4g}).

\begin{lstlisting}[xleftmargin=5.0ex,language=MyJava, caption={A GUI listener defined as a class}, label=lst.listClass]
class IconPaneMouseListener implements MouseListener {
  %%@Override%% public void mouseClicked(MouseEvent e) {|\label{code.0g}|
    if(!isDrag) {|\label{code.3g}|
      //...
    }
  }
  %%@Override%% public void mouseReleased(MouseEvent e) {|\label{code.1g}|
    isDrag = false;|\label{code.4g}|
  }
  %%@Override%% public void mouseEntered(MouseEvent e) {|\label{code.2g}|
    isMouseExited = false; 
    // ...
}}
\end{lstlisting}

\section{Related Work}\label{sec.related}

Work related to this paper fall into two categories:
design smell detection;
GUI maintenance and evolution.

\subsection{Design Smell Detection}

\change{The characterization and detection of object-oriented (OO) design smells have been widely studied~\cite{rasool2015review}. 
For instance, research works characterized various OO design smells associated with code refactoring operations~\cite{fowler1999,brown1998}.
Multiple empirical studies have been conducted to observe the impact of several OO design smells on the code.
These studies show that OO design smells can have a negative impact on maintainability~\cite{Yamashita2013}, understandability~\cite{Abbes2011}, and change- or fault-proneness~\cite{Khomh2012}.
While developing seminal advances on OO design smells, these research works focus on OO concerns only.
Improving the validation and maintenance of GUI code implies a research focus on GUI design smells, as we propose in this paper.}

%
%

\change{Related to GUI code analysis, Silva \etal propose an approach to inspect GUI source code as a reverse engineering process~\cite{silva2010gui,Silva2010}.
Their goal is to provide developers with a framework supporting the development of GUI metrics and code analyzes.
They also applied standard OO code metrics on GUI code~\cite{silva2014approach}.
Closely, Almeida \etal propose a first set of usability smells~\cite{Almeida2015}.
These works do not focus on GUI design smell and empirical evidences about their existence, unlike the work presented in this paper.}

\change{The automatic detection of design smells involves two steps.
First, a source code analysis is required to compute source code metrics. 
Second, heuristics are applied to detect design smells on the basis of the computed metrics to detect design smells.
Source code analyses can take various forms, notably: static, as we propose, and historical.
Regarding historical analysis, Palomba \etal propose an approach to detect design smells based on change history information~\cite{palomba14}.
Future work may also investigate whether analyzing code changes over time can help in characterizing \bls.
Regarding detection heuristics, the use of code metrics to define detection rules is a mainstream technique.
Metrics can be assemble with threshold values defined empirically to form detection rules~\cite{moha2010}. 
Search-based techniques are also used to exploit OO code metrics~\cite{Sahin14}, as well as machine learning~\cite{Zanoni2015}, or bayesian networks~\cite{khomh2011}.
Still, these works do not cover GUI design smells.
In this paper, we focus on static code analysis to detect GUI commands to form a \bl detection rule.
To do so, we use a Java source code analysis framework that permits the creation of specific code analyzers~\cite{spoon}.
Future work may investigate other heuristics and analyses to detect GUI design smells.
}

Several research work on design smell characterization and detection are domain-specific.
For instance, Moha~\etal propose a characterization and a detection process of service-oriented architecture anti-patterns~\cite{moha12}.
Garcia \etal propose an approach for identifying architectural design smells~\cite{garcia2009}.
Similarly, this work aims at motivating that GUIs form another domain concerned by specific design smells that have to be characterized.

Research studies have been conducted to evaluate the impact of design smells on system's quality~\cite{olbrich2010,fontana2013} or how they are perceived by developers~\cite{palomba2014}.
\change{Future work may focus on how software developers perceive \bls.}

\subsection{GUI maintenance and evolution}

Unlike object-oriented design smells, less research work focuses on GUI design smells.
Zhang \etal{} propose a technique to automatically repair broken workflows in Swing GUIs~\cite{ZhangLE2013}.
Static analyses are proposed.
This work highlights the difficulty "\emph{for a static analysis to distinguish UI actions [GUI commands] that share the same event handler [GUI listener]}".
In our work, we propose an approach to accurately detect GUI commands that compose GUI listeners.
Staiger also proposes a static analysis to extract GUI code, widgets, and their hierarchies in C/C++ software systems~\cite{Staiger07}.
The approach, however, is limited to find relationships between GUI elements and thus does not analyze GUI controllers and their listeners.
Zhang \etal{} propose a static analysis to find violations in GUIs~\cite{Zhang2012}.
These violations occur when GUI operations are invoked by non-UI threads leading a GUI error.
The static analysis is applied to infer a static call graph and check the violations.
Frolin \etal{} propose an approach to automatically find inconsistencies in MVC JavaScript applications~\cite{Frolin15}.
GUI controllers are statically analyzed to identify consistency issues (\eg inconsistencies between variables and controller functions).
This work is highly motivated by the weakly-typed nature of Javascript.



\section{Conclusion}\label{sec.conclu}

\change{In this paper, we investigate a new research area on GUI design smells.
We detail a specific GUI design smell, we call \bl, that can affect GUI listeners.
The empirical study we conducted exhibits a specific number of GUI commands per GUI listener that characterizes a \bl exists.
We define this threshold to three GUI commands per GUI listener.
We show that \SI{21}{\percent} of the analyzed GUI controllers are affected by \bls.
We propose an algorithm to automatically detect \bls.
This algorithm has been implemented in a tool publicly available and then evaluated.

Next steps of this work include a behavior-preserving algorithm to refactor detected \bls.
We will conduct a larger empirical study to investigate more in depth the relation between the number of bug fixes over the number of GUI commands.
We will study different GUI coding practices to identify other GUI design smells.
We will investigate whether some GUI faults~\cite{LEL15} are accentuated by GUI design smells.}

\section*{Acknowledgements}
This work is partially supported by the French BGLE Project CONNEXION.
We thank Yann-Ga\"el Gu\'eh\'eneuc for his insightful comments on this paper.

\clearpage

\balance

\bibliographystyle{SIGCHI-Reference-Format}
\bibliography{ref}

\end{document}